\documentclass[lettersize,journal]{IEEEtran}
\usepackage{amsmath,amsfonts}
\usepackage{algorithmic}
\usepackage{algorithm}
\usepackage{array}
\usepackage[caption=false,font=normalsize,labelfont=sf,textfont=sf]{subfig}
\usepackage{textcomp}
\usepackage{stfloats}
\usepackage{url}
\usepackage{verbatim}
\usepackage{graphicx}
\usepackage{cite}

\begin{document}

\title{Weighting Factors Tuning by Direct Feedback in Predictive Control of Multiphase Motors}

\author{Manuel R. Arahal$^1$, Manuel G. Satué$^1$, Kumars Rouzbehi$^1$, Francisco Colodro$^2$
\thanks{$^1$Department of Systems Engineering and Automation, Universidad de Sevilla, Seville, Spain.}

\thanks{$^2$Department of Electronic Engineering, Universidad de Sevilla, Seville, Spain.}

}

\maketitle

\begin{abstract}
Predictive Stator Current Control (PSCC) has been proposed for control of  multi-phase drives. The flexibility offered by the use of a Cost Function has been used to deal with the increased number of phases. However, tuning of the Weighting Factors constitutes a problem. Intensive trial and error tests are usual in this context. Existing on-line selection methods, on the other hand, require  large amounts of data and/or complex optimization procedures. The proposal of this paper is a closed-loop scheme that links Weighting Factors to performance indicators. In this way, optimal Weighting Factors are determined for each operating point. Also, changes in reference values for performance indicators are easily tackled. Unlike previous  methods, the proposal carries very little computational burden. A case study is developed for a five-phase induction motor and assessed with real experimentation on a laboratory set-up.
\end{abstract}


\section{Introduction}
Predictive Stator Current Control (PSCC) is a form of Model Predictive Control (MPC) for drives. The method follows the scheme of Finite State MPC (FSMPC), where the Voltage Source Inverter (VSI) states are directly commanded. PSCC has been used in the context of multi-phase systems due to its flexibility to incorporate different control objectives \cite{tenconi2018model}. This is done including different terms in the Cost Function (CF) that is minimized every sampling period. Weighting Factors (WF) are used as coefficients in the CF to assign more or less importance to each term \cite{zerdali2024review}. For instance, in Induction Machines (IM), stator current errors in $\alpha-\beta$ plane is an usual term that must be balanced with stator current error in $x-y$ plane and with Average Switching Frequency (ASF) \cite{wang2022model}.

Early papers of PSCC for multi-phase systems recognized the problem of CF tuning and explored different tunings as well as controller variations \cite{liu2017overview}. The usual practice was testing different CF tuning until a particular compromise solution was found. Later on it was realized that, each operating point required a specific CF tuning to meet the expected behavior. Some research effort was directed to deal with this. For instance, in \cite{gonzalez2019constrained} soft constraints are introduced in the predictive formulation for a nine-phase IM drive and the effect of cost function parameters is discussed. Partition of the operating space was proposed in \cite{arahal2019cost} as a means of providing different tunings for different operating regions. This idea has also appeared in related applications such as multilevel rectifiers \cite{makhamreh2020lyapunov}. 

Another aspect of the problem of CF tuning is that the usual performance indicators for PSCC show trade-offs. This was first shown in a systematic way in \cite{arahal2018trade} and later also recognized in several works such as \cite{fretes2021pareto}, where a Pareto frontier is presented for a 6-phase machine. These links between figures of merits have been quantitatively described in \cite{arahal2021adaptive}, where a simple cubic approximation is derived for the case of PSCC of a a five-phase IM. This result allows the use of a simplified adaptation rule to incrementally tune the CF on-line. 

\subsection{Critical literature review}

The first known method for systematic and automated WF tuning was proposed for a  Shunt Active Power Filter in \cite{zanchetta2011heuristic}, where a genetic algorithm was used. The Pareto front is not characterized mathematically. This lack of insight makes the derivation of the WF unnecessarily complex, necessitating a general optimization method. Other works use similar ideas. Comparing with \cite{zanchetta2011heuristic}, changes are found in the number of figures of merit, and the search procedure. All  these works share the criticisms made for \cite{zanchetta2011heuristic}.

Systematic WF tuning for VSD was first tackled in  \cite{arahal2018trade} for PSCC of a five-phase IM using  the Pareto front for two WF. In  \cite{guazzelli2018weighting} WF selection is performed for Predictive Torque Control (PTC) of a six-phase IM using Multi Objective Genetic Algorithm. However, the $x-y$ plane issue is not considered. NSGA-II With TOPSIS Decision Making is used in \cite{arshad2019weighting} for PTC of a conventional IM. The dependence on operating point is restricted to load and no-load conditions. In \cite{fretes2021pareto} a multi-objective particle swarm optimization is used. Both $x-y$ currents and switching frequency are considered for a six phase IM, confirming the results of \cite{arahal2018trade}. In \cite{yao2022ann}, artificial neural network is used for two WF in a permanent magnet drive. In all these works the WF analysis uses simulation later supported by experimentation on a few cases. Thus, these proposals belong to the off-line class.

Regarding on-line methods for VSD, the work of \cite{muddineni2017finite} presents, for the first time, a variable WF method for PTC of an IM. It also demonstrates non-linear variations of performance indicators vs. speed (see Fig. 13 in \cite{muddineni2017finite}). However, the WF tuning is made at every sampling period although the effect cannot be sensed immediately. In \cite{ravi2018enhanced} PTC of a three-phase IM uses a single optimal WF found at every sampling period. However, the method does not link actual figures of merit to WF.

In contrast with the above, in \cite{arahal2021adaptive} an adaptive cost function is proposed linking performance to WF. The VSD used is a six-phase IM with two WF in the CF. The method clearly links figures of merit to WF (see Fig. 5 in \cite{arahal2021adaptive}). The Pareto front is characterized by a single equation, allowing simple rules to be used for adaptation. In this way, the heftiness of general methods (such as genetic, swarm and neural) is avoided. However, the proposal needs more testing and the adaptation mechanism has to be made more resilient to irregularities in the derivatives.

More recently, some works have dealt with on-line WF tuning of various VSD. In most cases, cumbersome algorithms are used instead of simple rules. This is the case of \cite{deng2024reinforcement}, where the artificial intelligence technique known as reinforcement learning is used to tune the WF of PTC for a permanent magnet motor. This technique is know for being difficult to tune. It also requires a large amount of data and computation. Similar criticism can be made for other, similar, works. Nevertheless, the work presented in \cite{shahid2024optimal} is interesting, as they clearly show the effect of speed and torque on figures of merit. A Multi-Criteria-Decision-Making technique relying on a data set of the control objectives is used to derive WF for PTC of an IM.

\subsection{Novelty and contributions}
This paper continues with the line of work of \cite{arahal2021adaptive} by proposing a new on-line WF selection method. In contrast with previous proposals, the method is quite simple, does not require data-base or complex swarm/distributed optimization. Also, and more importantly, the proposal is a closed-loop method. This means that the figures of merit are monitored and their value used to drive the WF change. Despite the simplicity of this approach, it has not been proposed before. Hence, the novelty of the proposal lies in the fact that the on-line WF selection has been solved by means of a very simple solution based on feed-back. 

The contributions of the paper are as follows.
\begin{enumerate}
    \item A novel adaptive procedure is proposed for on-line tuning of WF of multi-phase PSCC. In this way optimal values of WF are found without resorting to off-line trial and error procedure.
    \item The procedure grants automatic adaptation of WF to the actual operating regime of the motor.
    \item The procedure can be used for situations in which the desired values of performance indicators change for whatever reason.
    \item The procedure is tested in a real induction motor.
\end{enumerate}

The proposal is assessed with real experimentation in a five-phase drive IM. The proposal can be used with other systems, such as power converters supplying a static electrical load or with other types of motors.

The rest of the paper is organized as follows, Section 2 summarizes the PSCC for a five-phase IM and introduces the figures of merit considered. Next, in Section 3, the new method is presented. Section 4 is devoted to assessing the proposal using simulation and experimental results for a laboratory setup using a five-phase IM.

\section{Predictive stator current control}
\begin{figure}
\centering
\includegraphics[width=8cm]{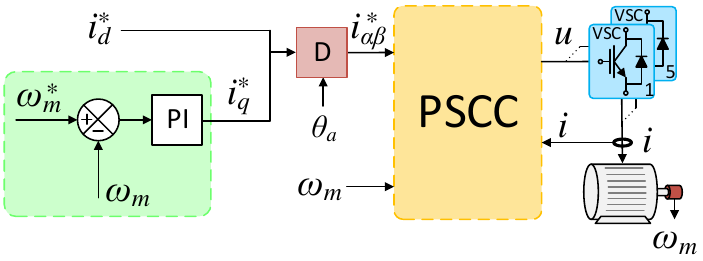}
\caption{\label{f_diagr_IFOC} Diagram of the PSCC method.}
\end{figure}

The control of stator currents in multi-phase motors is based on 
Indirect Field-Oriented Control (IFOC). In  the IFOC scheme, flux and  torque are independently regulated. The control scheme is given in Fig. \ref{f_diagr_IFOC}. The stator flux current set point $i^{*}_{ds}$ is set to magnetize the motor whereas quadrature current $i^{*}_{qs}$ is used to manipulate the produced torque. The reference value of the magnetizing current, $i^{*}_{ds}$, is kept constant, and the measured error is fed into a PI controller which computes the reference for the direct component of the stator voltage, $v^{*}_{ds}$.

The speed controller (another PI) in the velocity feedback loop is responsible for generating $i^{*}_{qs}$ to drive the mechanical speed control error to zero. 
\begin{align}\label{eq_pi_veloc}
i^{*}_{qs}(t) = k_p \cdot e(t) + k_i \int_0^t e( \tau) d \tau
\end{align}
\noindent where $e(t) = \omega^*(t) - \omega^e(t)$ is the velocity error or difference between the speed set point ($\omega^*$) and the speed measurement ($\omega^e$).

Once the set-points in $d-q$ coordinates are known, they are projected to the $\alpha-\beta$ space using the Park transformation, obtaining a reference for stator current in $\alpha-\beta$ plane as $ I^*_{\alpha-\beta} = D {\left( i^{*}_{ds}, i^{*}_{qs} \right)}^{\mathrm{t}} $, where matrix $D$ is given by 
\begin{align}\label{eq_park}
  D = 
  \begin{pmatrix}
   ~\cos ~ \theta_a  & \sin ~ \theta_a \\
   -\sin ~ \theta_a &  \cos ~ \theta_a\\
  \end{pmatrix}
\end{align}

The flux position $\theta_a$ is estimated as $\theta_a = \int \omega_e \, dt$ where $\omega_e = \omega_{sl} + P \omega$, being $P$ the number of pairs of poles of the IM and
\begin{equation}\label{eq_slip}
  \omega_{sl} = \frac{i^{*}_{qs}}{i^{*}_{ds}} \frac{1}{\hat{\tau}_r}
\end{equation}

\noindent where $\hat{\tau}_r$ is an estimation of the rotor time constant $\tau_r = L_r/R_r$, being $L_r$ the rotor inductance and $R_r$ the rotor resistance, respectively. As a result, the set point for stator current tracking $i^*(k)$ has an amplitude $I^* = \sqrt{  i^{*2}_{ds} +  i^{*2}_{qs} }$. Finally, the $\alpha-\beta$ references can be expressed as $i^*_{\alpha}(t)=I^* \sin{ \omega_e t}$, $i^*_{\beta}(t)=I^* \cos{ \omega_e t}$, $i^*_{x}(t)=0$, $i^*_{y}(t)=0$.

Tracking of stator currents references $i^*_s$ is the job of the inner loop controller. In FSMPC, a predictive model is used to predict stator currents in $\alpha-\beta$ and $x-y$ planes for discrete time $(k+2)$. This two-step procedure is needed to compensate for the digital delay caused by computations \cite{arahal2018trade}.

At discrete time $(k)$, the VSI state that provides a lower value for a cost function is selected. This state is defined by vector  $u = \left( K_a, K_b,  \cdots , K_e  \right)^\top$, where each $K_{ph}$ value indicates the state of the corresponding switch for phases $ph = \{ a, b, c, d, e \}$, in the five-phase VSI. The optimized value $u^o$ is then issued for the whole $(k+1)$ period. At the end of this period the whole procedure is repeated.

The predictive model assumes the form of discrete-time state-space equations of the form 
\begin{equation}\label{eq_p1p}
 \hat{i}_s(k+1) = \Phi ( \omega) i_s(k) +  \Psi u(k),
\end{equation}

\noindent where where $i_s$ are stator currents in $\alpha-\beta$ and $x-y$ planes, and $\omega$ is the angular speed. Matrices $\Phi$ and $\Psi$ are obtained from basic IM modelling \cite{kindl2020spatial}. The prediction for $(k+2)$ is found as
\begin{equation}\label{eq_p2p}
 \hat{i}_s(k+2) = \Phi( \omega) \hat{i}_s(k+1) + \Psi u(k+1) + G(k)
\end{equation}

\noindent where vector $G$ accounts for effect of rotor currents that are usually not measured. This term is computed by backtracking as $G(k) = i_s(k) - \hat{i}_s(k)$.

The cost function imposes penalties on $\alpha-\beta$ tracking error, $x-y$ content, and instantaneous number of commutations in the VSI legs. These terms have different weighting factors  ($\lambda$) to allow for different tuning options. The mathematical expression of the cost function is 
\begin{equation}\label{eq_fcoste_NC}
  J(k+2) = \| \hat{e}_{\alpha\beta} (k+2) \|^2 + \lambda_{xy} \| \hat{e}_{xy}(k+2) \|^2 +  \lambda_{sc} SC(k+1),
\end{equation}

\noindent where $\| . \|$ denotes vector modulus, $\hat{e} =  \left( i^*_s - \hat{i}_s \right)$ is the predicted current error, and $SC$ stands for Switch Changes. The $SC$ is computed for a VSI change from $u(k)$ to $u(k+1)$ as
\begin{equation}\label{eq_def_NC}
  SC(k+1) = \sum_{ph=1}^5 | K_{ph}(k+1) - K_{ph}(k) |.
\end{equation}

\begin{figure}
\centering
  \includegraphics[width=9cm]{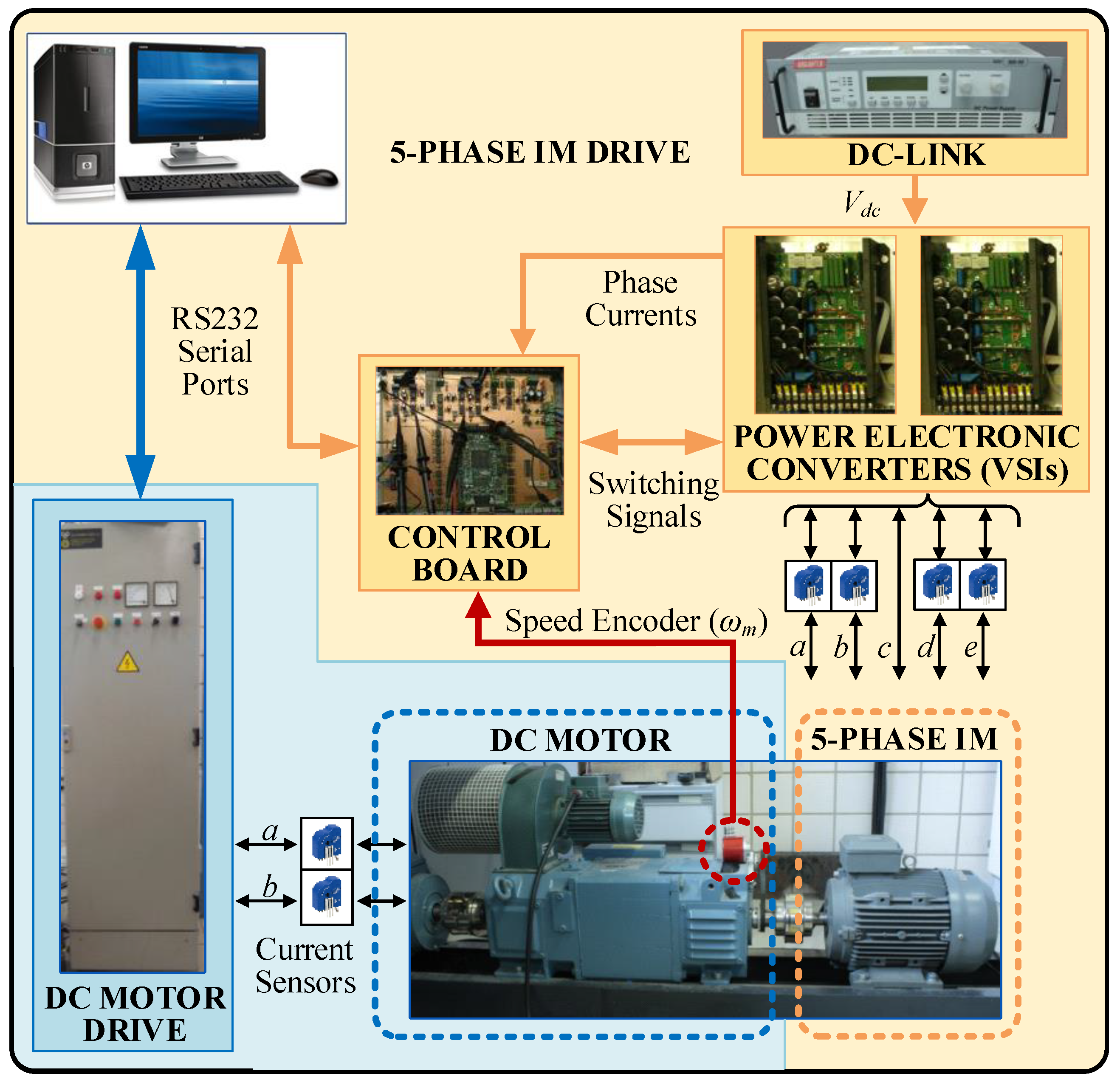}\\
  \caption{Elements of the experimental 5-phase IM drive.}\label{fig_bancada}
\end{figure}

\subsection{Laboratory Setup}
A laboratory test-bench is used to validate the proposal. The setup allows experimentation on a real five-phase induction machine as shown in Fig. \ref{fig_bancada}. The experimental system includes an induction motor with five phases and  parameters shown in Table \ref{tab_Parameters}. The VSI uses two three-phase SEMIKRON SKS 22F modules powered by a 300V DC power supply. The control uses a MSK28335 board including a TMS320F28335 Digital Signal Processor. The control program is run in real-time with sampling period $T_s$ in the order of 30 ($\mu$s) depending on the computational load of the controller.

\begin{table}
\centering
\caption{Parameters of the Experimental 5-phase IM.\label{tab_Parameters}}
\begin{tabular}{ccc}
 \textbf{ Parameter} & \textbf{ Value} & \textbf{ Units}  \\\hline
  Stator resistance, $R_s$           &   12.85 &   $\Omega$  \\
 Rotor resistance, $R_r$            &	4.80 &   $\Omega$  \\ 
 Stator leakage inductance, $L_{ls}$&   79.93 &   mH \\
 Rotor leakage inductance, $L_{lr}$ &   79.93 &   mH \\
 Mutual inductance, $L_M$           &  681.7  &   mH \\
 Rotational inertia, $J_m$          &    0.02 &   kg m$^2$ \\
 Number of pairs of poles, $P$      &  3      & - \\ \hline
\end{tabular}
\end{table}

\subsection{Performance indices and weighting factors}
The goal of the inner loop is the generation of adequate stator currents with a proper use of the VSI. The assessment of FSMPC controllers often use the following performance indices
\begin{eqnarray} \label{eq_gamma1}
  \Gamma_1 & =&  \sqrt{ \frac{1}{N} \sum_{k=1}^{N} e_{\alpha\beta}^2 (k) } \\ \label{eq_gamma2}
  \Gamma_2 &=& \sqrt{ \frac{1}{N} \sum_{k=1}^{N} e_{xy}^2 (k) },
\end{eqnarray}

\noindent where $N$ is the number of sampling periods considered for the computation. Please notice that these $\Gamma$ values are root mean squared control errors in $\alpha-\beta$ and $x-y$ planes respectively.

A third index is needed because in PSCC the VSI switching frequency is not constant. However, the Average Switching Frequency is more or less constant for a particular operating regime of the motor. The ASF has a relevant impact on the selection of the hardware (e.g. standard IGBTs or SiC-based power switches) and on efficiency (VSI losses). The number of commutations in the VSI can be used to estimate the ASF as 
\begin{equation}\label{eq_gamma3}
  \Gamma_3 = \frac{2 \pi}{N \cdot T_s \cdot \omega_e} \sum_{k=1}^{N} \frac{SC(k)}{6},
\end{equation}

\noindent where $T_s$ is the sampling period.

It has been pointed out in previous papers that, by carefully tuning of the WF, the  figures of merit can be tuned. It must be clear that, an increase in $\lambda_{xy}$ should produce a reduction in $\Gamma_2$. Similarly, an increase in $\lambda_{sc}$ should produce a reduction in $\Gamma_3$. However, this simple relationships are complicated by several facts.

\begin{enumerate}
    \item Links between figures of merit make any WF to affect all $\Gamma$ to some extent.
    \item The relationship between WF and $\Gamma$ depend on the operating regime of the drive.
    \item Improvement of all $\Gamma$ at once by tuning the WF is impossible past the Pareto front \cite{arahal2018trade, fretes2021pareto}.
\end{enumerate}

As a result, WF tuning  always means trading some indicators in exchange for others. The proposed scheme aims at doing this kind of WF changes on-line and in closed-loop.

\section{Closed loop tuning of WF}
\begin{figure}
\centering
  \includegraphics[width=8.5cm]{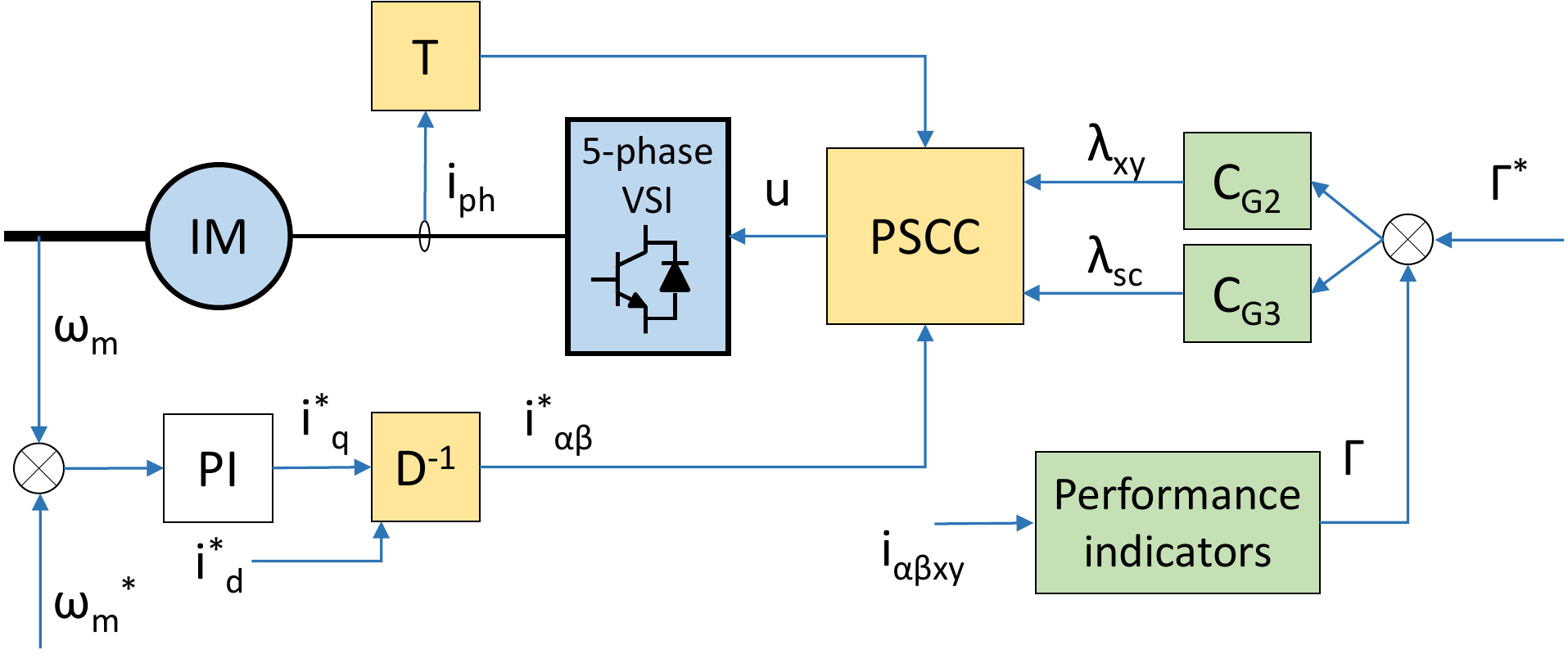}\\
  \caption{Diagram of the proposal for closed-loop tuning of WF.}\label{fig_diagr_bgc}
\end{figure}

The proposal uses a closed loop scheme to set the values of the WF during the normal operation of the motor as indicated in Fig. \ref{fig_diagr_bgc}. In this way, the vector of weighting factors $\Lambda = (\lambda_{xy}, \lambda_{sc})$ can be determined for each operating regime of the motor and for each desired value of the performance indicators ($\Gamma^*$). Two feed-back controllers $C_{G2}$ and $C_{G3}$ are used for this task. The first one uses $\lambda_{xy}$ as manipulated variable to control $\Gamma_2$ (i.e. drive $\Gamma_2$ to its reference value $\Gamma_2^*$). Similarly, $C_{G3}$ uses  $\lambda_{sc}$ as manipulated variable to control $\Gamma_3$.

To tune the controllers it is necessary to characterize the dynamics of the system. A step-test can be used to identify the main characteristics of said dynamics. To perform the test, an abrupt change is performed on the WF vector $\Lambda$. The evolution of stator currents is recorded. Using these measurements and the definitions of equations (\ref{eq_gamma1})-(\ref{eq_gamma3}), the 
$\Gamma$ values are obtained.

\begin{figure*}[tp]
  \centering
  \begin{tabular}{cc}
  \includegraphics[width=8cm]{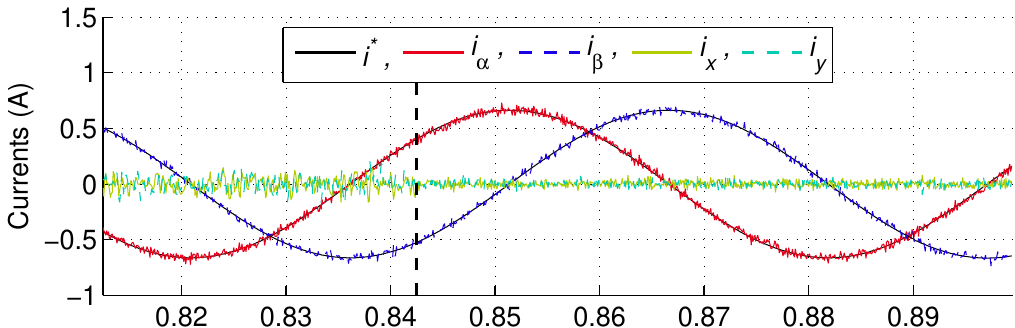} & 
  \includegraphics[width=8cm]{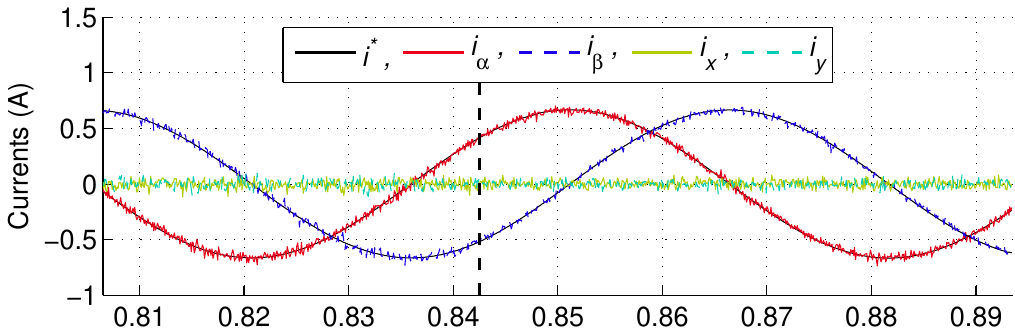} \\
  \includegraphics[width=8cm]{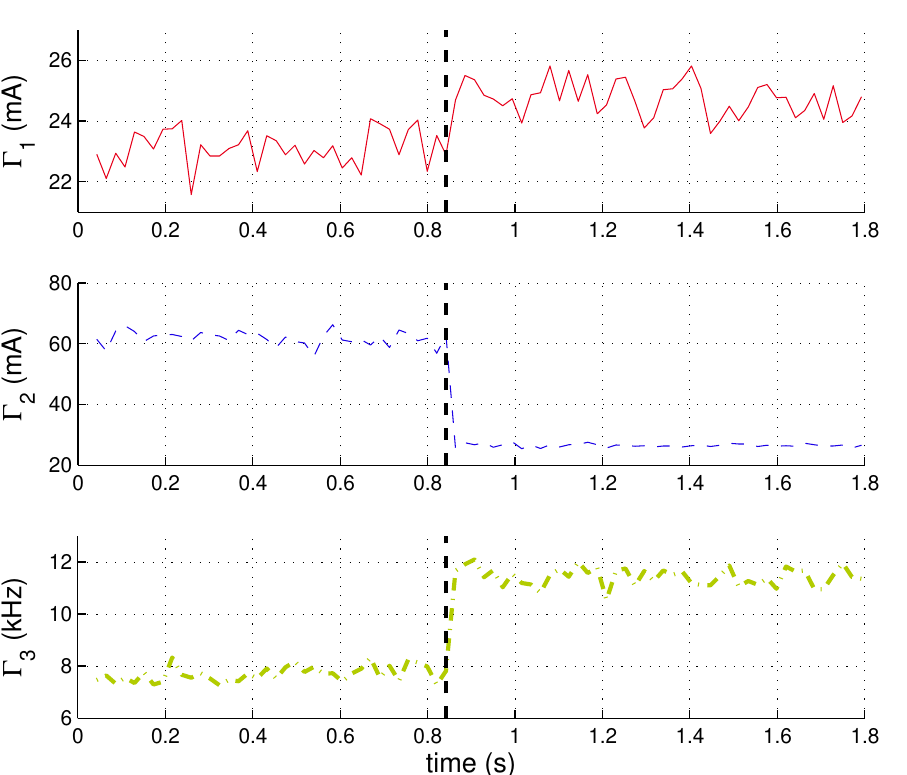} & 
  \includegraphics[width=8cm]{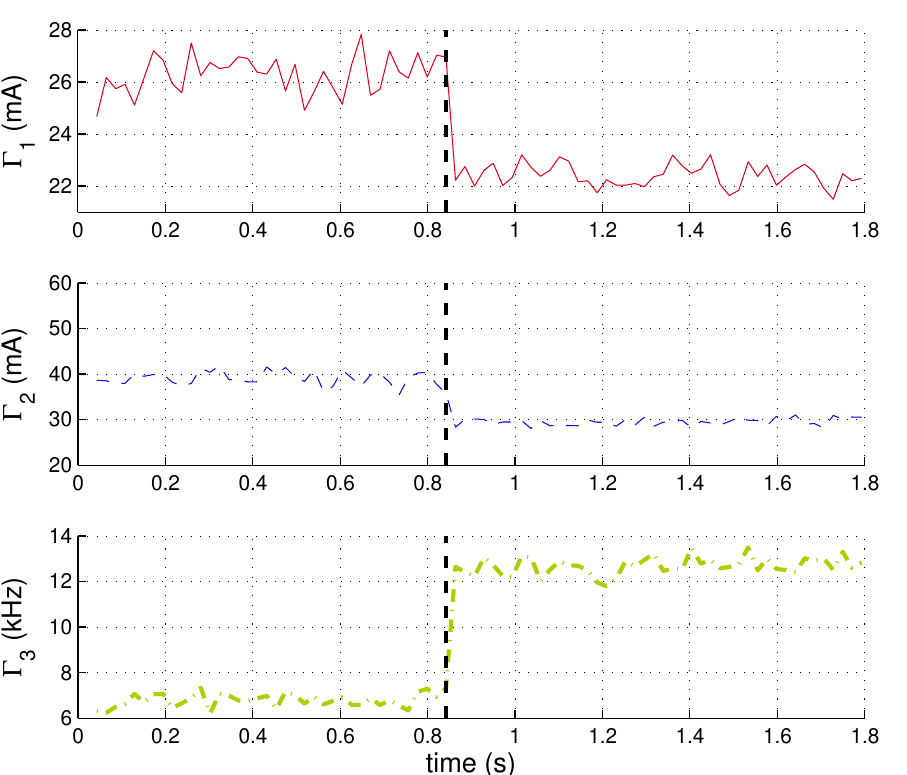} \\
  \end{tabular}
  \caption{Evolution of electrical variables for two  WF step change tests. Notice the different time scale in top and bottom graphs.}\label{fig_stepWF1}
\end{figure*}

As an example consider a change from $\Lambda = (0.15, 20 \cdot 10^{-4})$ to $\Lambda = (0.75, 20 \cdot 10^{-4})$ (i.e. an increase in $\lambda_{xy}$ with $\lambda_{sc}$ constant). With this change a reduction in $\Gamma_2$ is expected. Also, the other figures of merit should increase as a result of the trade-offs already discussed. The waveforms in the left pane of Fig. \ref{fig_stepWF1} show the stator currents in $\alpha-\beta$ and $x-y$ axes around the time of change (indicated with a vertical dashed line). The WF change causes a visible change in the behavior of stator currents.  As expected, $x-y$ currents are greatly reduced. More importantly, the adaptation of currents is almost immediate. This is due to the fact that PSCC is a memory-less sub-system: i.e. the optimization of the control action is not influenced by past events. In this way, changes in WF have the effect of changing the cost function optimization results immediately. 

The trajectories of $\Gamma$ values are also shown in the left pane of Fig. \ref{fig_stepWF1}. The changes in performance indicators are as expected given the links between figures of merit. Also, the quick response to WF changes is clearly visible. This might led to the conclusion that changing the WF at every sampling period is positive. This is a naive conclusion because the performance indicators cannot be measured instantaneously. Recall that $N$ samples are needed to record a new value for any of the $\Gamma$ values. A value $N=720$ (samples) is used for this test. Notice that, using this $N$ and for a sampling period of $T_s=30$ ($\mu$s), the $\Gamma$ values are updated every $0.022$ seconds. This value is acceptable as changing in operating regimes takes considerably longer.

The right pane of Fig. \ref{fig_stepWF1} show the results for another WF change. In this case an initial value of $\Lambda = (0.4, 26 \cdot 10^{-4})$ is changed mid-time to $\Lambda = (0.4, 14 \cdot 10^{-4})$ (i.e. a reduction in $\lambda_{sc}$ with $\lambda_{xy}$ constant). With this change an increase in $\Gamma_3$ is expected.

These two step test also illustrate the fact that each WF affects all figures of merits as indicated in \cite{arahal2018trade}. This makes the design of a closed loop WF method a bit more complicated as will be shown next.

\subsection{Controller structure and tuning}
A simple PI structure is used for both $C_{G2}$ and $C_{G3}$ controllers. This structure needs specification of the reference value for $\Gamma_2$ and $\Gamma_3$. These values are application-dependent and can also be made operating-point-dependent. Denoting the reference value as $\Gamma_h^*$ (for $h=2,3$) it is possible to write 
\begin{align}\label{eq_pi_gamma}
  \Lambda_h (t) = g_{p,h} \cdot e_{\Gamma,h} (t) + g_{i,h} \int_0^t e_{\Gamma,h}( \tau) d \tau
\end{align}

\noindent where $\Lambda_h$ stands for the components of the WF vector, with $\Lambda_2=\lambda_{xy}$ and $\Lambda_3=\lambda_{sc}$. Also, $e_{\Gamma,h} = \Gamma^*_h  - \Gamma_h$ is the difference between the desired value for the performance indicator and its actual value. Finally, coefficients $g_{p,h}$ and $g_{i,h}$ are the proportional and integral gains of the PI blocks.

Tuning of $C_{G2}$ and $C_{G3}$ is done selecting values for $g_{p,h}$ and $g_{i,h}$ for $h=2,3$. Many different procedures have been proposed in the past for tuning of this kind of controllers. The reader is referred to \cite{skogestad2001probably} for an account of simple rules. For this paper, and as a proof of concept, tuning is based on said simple concepts. This results in the following gains that will be used in the next tests: $g_{p,2}=-1$, $g_{i,2}=-2.8$, $g_{p,3}=-4.5 \cdot 10^{-8}
$, $g_{i,3}=-10^{-7}$.

To assess the WF adaptation, some tests are performed in which the objectives (i.e. references for the performance indicators) suffer a step change. In a practical situation these changes can be useful to address different operating modes. As an example consider the case where one wishes to balance tracking performance with energy efficiency. Since commutations and $x-y$ content are sources of energy inefficiency, one might be willing to trade some $\Gamma_1$ in order to reduce both $\Gamma_2$ and $\Gamma_3$. This might be of use in different applications for instance electric vehicles driving on performance vs. economic mode.

For the first test, $\Gamma_2^*$ is changed from $\Gamma_2^*=50$ (mA)  to $\Gamma_2^*=30$ (mA). As a result, controller $C_{G2}$ should issue a new value for $\lambda_{xy}$. But this can cause a change in $\Gamma_3$ that constitutes a disturbance for controller $C_{G3}$. As a result a change in  $\lambda_{sc}$ is to be expected. The value of $\Gamma_1$ will settle in its corresponding value, lying in the Titeica surface described in \cite{arahal2021adaptive}.

\begin{figure*}[tp]
  \centering
  \begin{tabular}{cc}
  \includegraphics[width=8cm]{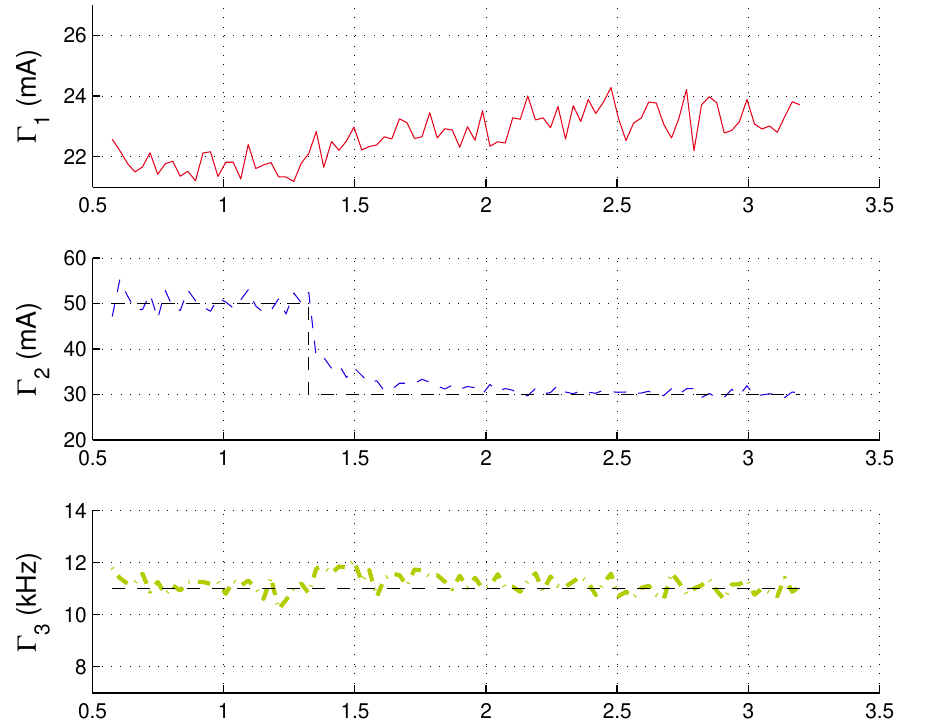} & 
  \includegraphics[width=8cm]{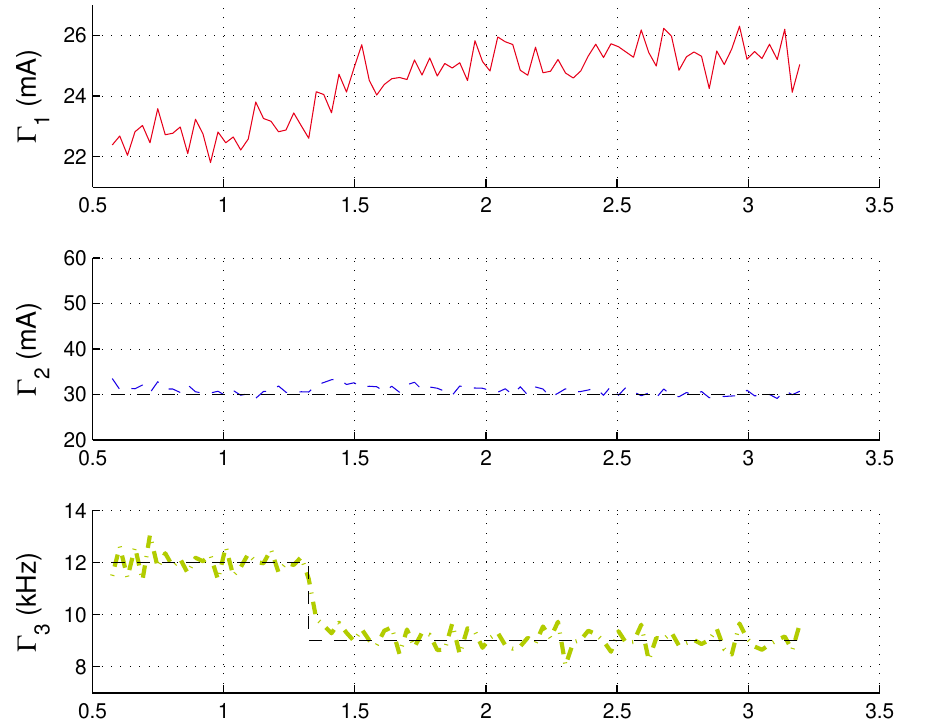} \\
  \includegraphics[width=8cm]{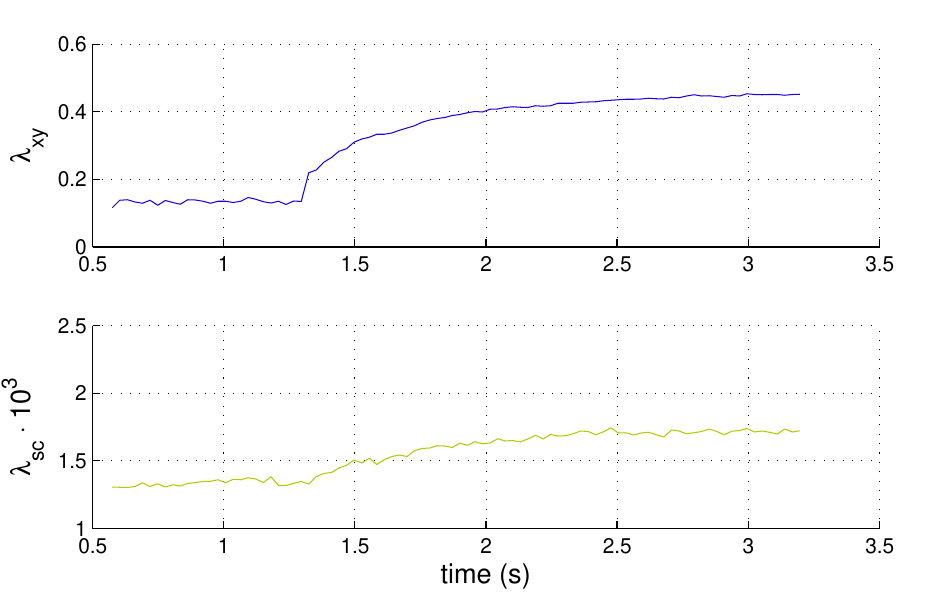} & 
  \includegraphics[width=8cm]{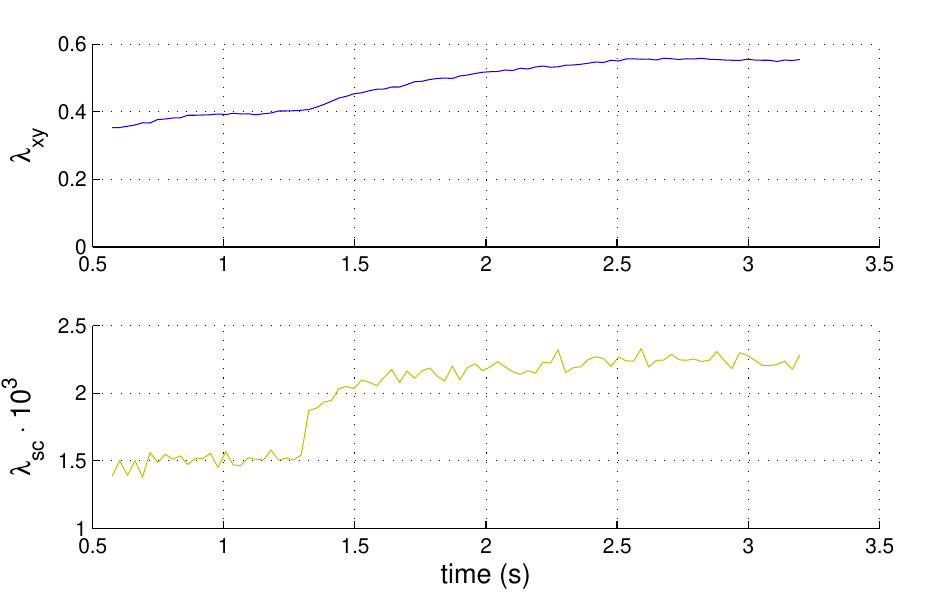} \\
  \end{tabular}
  \caption{Evolution of performance indicators (top) and weighting factors (bottom) for two step tests.}\label{fig_stepGref}
\end{figure*}

The results shown in the left pane of Fig. \ref{fig_stepGref} show the evolution of $\Gamma$ values as a result of the change in $\Gamma_2^*$. The adaptation of WF values can be observed. The trajectories of both $\lambda_{xy}$ and $\lambda_{sc}$ exhibit a well damped behavior. The closed loop characteristic time less than 1 second, which is fast enough compared with the mechanical time constant of the system.

In a second test, $\Gamma_3^*$ is changed from $\Gamma_3^*=12$ (kHz)  to $\Gamma_3^*=9$ (kHz). In this case controller $C_{G3}$ must increase  $\lambda_{sc}$. This might disturb $\Gamma_2$, so a change in $\lambda_{xy}$ is expected. The results shown in the right pane of Fig. \ref{fig_stepGref} show the evolution of $\Gamma$ values as a result of the change in $\Gamma_3^*$. Again, the performance indicators follow their references.

\subsection{Mechanical transients}
\begin{figure}[htp]\centering
  \includegraphics[width=7cm]{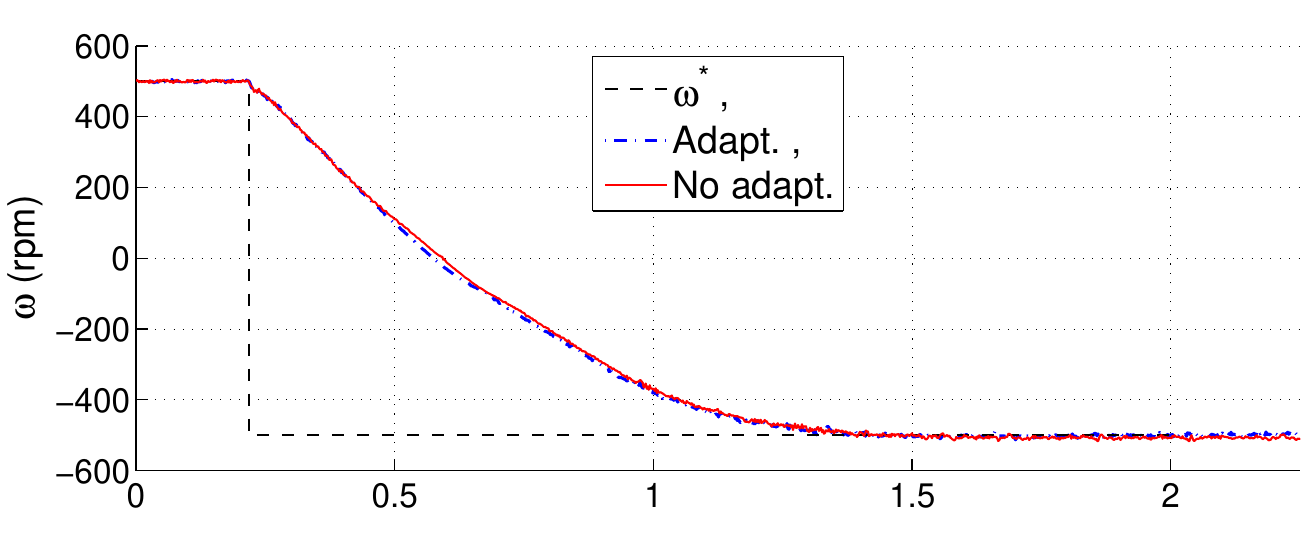}
  \includegraphics[width=7cm]{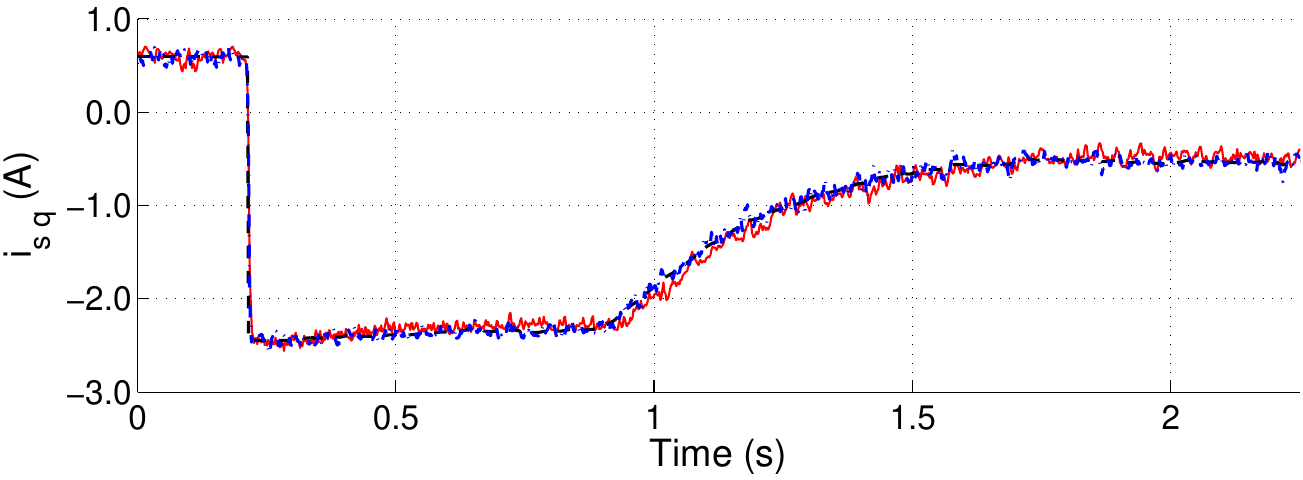}
  \caption{Results obtained in a reversal test with and without adaptation of the WF in terms of mechanical speed (left) and  $i_{sq}$ (right)}  
  \label{fig_result_reversal}
\end{figure}

The proposal provides also good results regarding  mechanical transients. It must be noted, however, that the performance in terms of mechanical speed is mainly set by the PI in the speed loop. For the comparison this PI is maintained in the same setting for the proposal and for the not adaptive case. The results are thus quite similar as shown in Fig. \ref{fig_result_reversal}.

\section{Conclusions}
Tuning of WF has received attention in the past as one open issue related to predictive control of power converters. The results obtained with   the proposal constitute an indication that simple adaptive tuning of WF is possible. 

The fact that the IM responds quickly in terms of electrical quantities indicates that fast WF adaptation is possible in principle. The results show, however, that sensing the effects of changes in WF is not immediate. This prompts for adaptation schemes that are parsimonious. This is in contrast with previous proposals.

Finally, the results of the paper confirm, once more, the existence of trade-offs between performance indicators. This fact is not highlighted in many works despite its importance.

\section{Acknowledgments}
This work is part of project I+D+i / PID2021-125189OB-I00, funded by MCIU/AEI/10.13039/ 501100011033/FEDER, UE ``ERDF A way of making Europe''.

\bibliographystyle{IEEEtran}
\bibliography{00main}

\vspace{11pt}

\vfill

\end{document}